\documentclass[twocolumn,showpacs,preprintnumbers,amsmath,amssymb]{revtex4}

\usepackage{graphicx}
\usepackage{dcolumn}
\usepackage{bm}
\usepackage[usenames]{color}
\usepackage{refcount}

\begin{document}

\preprint{APS/123-QED}

\title{Three-body recombination at vanishing scattering lengths in an ultracold Bose gas}

\author{Zav Shotan$^{1}\footnote{These authors contributed equally to this work}$}
\author{Olga Machtey$^{1*}$}

\author{Servaas Kokkelmans$^{2}$}%
\author{Lev Khaykovich$^{1}$}%
\affiliation{$^{1}$Department of Physics, Bar-Ilan University, Ramat-Gan, 52900 Israel}%
\affiliation{$^{2}$Eindhoven University of
Technology, P.O. Box 513, 5600 MB Eindhoven, The Netherlands}


\date{\today}

\begin{abstract}
We report on measurements of three-body recombination loss rates in an ultracold gas of $^7$Li atoms in the extremely nonuniversal regime where the two-body scattering length vanishes. We show that the loss rate coefficient is well defined and can be described by two-body parameters only: the scattering length $a$ and the effective range $R_e$. We find the rate to be energy independent, and, by connecting our results with previously reported measurements in the universal limit, we cover the behavior of the three-body recombination rate in the whole range from weak to strong two-body interactions. We identify a nontrivial magnetic field value in the nonuniversal regime where the rate should be suppressed.
\end{abstract}

\pacs{34.50.Cx, 21.45.-v, 67.85.-d}
\maketitle

The few-body problem underlies fundamental processes in physics, yet it is notoriously difficult for finding analytic and numerical solutions \cite{Faddeev}.
It challenges our mind with the complexity of small and system-dependent molecular structures at the size of their interaction potentials. 
At the same time, in the regime of resonant pairwise interactions, it provides an elegant description of unusually large bound states possessing universal properties.
Two important two-body length scales are involved in this description.
The first is the van der Waals length $r_{\textrm{vdW}}$, which is constant and connected to the radial range of the potential. 
The second is the $s$-wave scattering length $a$, which can be tuned magnetically via a Feshbach resonance~\citep{Chin10}. When $a\gg r_{\textrm{vdW}}$, universal two-body states with size $\sim a$ emerge, and a wealth of phenomena known as Efimov physics is opened up in three- and, generally, $N$-body sectors~\citep{Braaten&Hammer06,Wang13}. 
The fundamental states belonging to the Efimov effect (Efimov trimers) depend log-periodically on $a/r_{\textrm{vdW}}$~\citep{Wang12,Naidon12}.

Sufficiency of two-body parameters in the Efimov scenario indicates that two-body physics plays a decisive role in the universal few-body processes in ultracold gases.
The following questions arise as to how far this dominance of two-body physics extends when going to the opposite, nonuniversal limit, {\it i.e.} when $a\rightarrow 0$. 
What parameters govern the three-body processes and can we identify the influence of the short-range three-body forces in this limit? 
Although recent theoretical studies probe the non-universal regime~\citep{Petrov04,Lasinio10,Pricoupenko11}, their approaches cannot be directly extended to the regions of extreme nonuniversality, the scattering length zero crossings. 

Three-body recombination induced loss of atoms has been extensively used as an efficient probe of Efimov physics in trapped ultracold gases~\citep{Ferlaino11,Wild12,Rem13,Roy13,Huang14,Tung14,Pires14}.
The process involves collisions of three particles resulting in a formation of a two-body bound state and a free atom. 
The binding energy, usually much larger than the trap depth, is then released as kinetic energy of the colliding partners leading to losses.
In this Letter, we investigate three-body recombination rates in the vicinity of two different zero crossings in a gas of ultracold $^7$Li atoms. 
One zero crossing is associated with a broad Feshbach resonance, while the other one is connected to a narrow resonance. 
We observe a clear difference between the two regions with $a\approx 0$, and explain the magnetic field dependence of the three-body recombination loss rate coefficient in terms of an effective length parameter. 
This parameter describes the two-body scattering phase shift for vanishing scattering length, and can be understood in terms of the finite range of the two-body potential, given by the van der Waals length, and by the width of the Feshbach resonance.

Assuming the dominance of two-body physics in three-body processes, we start by considering the effective range expansion of the scattering phase shift $\delta(k)$ in its usual form~\citep{Taylor}:
\begin{equation}
k\cot\delta(k)=-\frac{1}{a}+\frac{R_e k^2}{2}.
\label{Eq_cotandelta}
\end{equation}
However, when $a\rightarrow 0$ the first term in the right hand of this equation diverges, and, to compensate this divergence, the second term diverges as well ($|R_e|\rightarrow\infty$), which makes the above expression inconvenient in this limit. 
A better form of the effective range expansion can be obtained by revamping Eq.~(\ref{Eq_cotandelta}) as 
\begin{equation}
-\frac{\tan\delta(k)}{k}=a - V_e k^2,
\label{Eq_tandelta}
\end{equation}
where $V_e=-R_e a^2/2$ is the effective volume~\cite{Blackley14}. 
We show that although neither $a$ nor $R_e$ are good lengths at zero crossings, their combination in the form of the effective length $L_e=V_e^{1/3}$ remains finite, and captures the behavior of the three-body recombination loss rate coefficient remarkably well~\footnote{For an isolated Feshbach resonance, it can be shown that $V_e=(R^* a^2_{bg} -r_{\textrm{vdW}}^3/3)$ exactly when $a=0$ \cite{Kokkelmans14}, where $a_{bg}$ is the background scattering length and $R^*$ is inversely proportional to the width of the resonance. 
However, for overlapping resonances, as is the case in $^7$Li, the analytic expression for the effective length is more involved and will be the subject of a future publication. Instead, we utilize, here, numerical values for $L_e$ resulting from a coupled-channels calculation.}.
Once $a$ and $L_e$ are known, no further information on the short-range two-body or three-body potentials is needed. 
Moreover, while the two-body collisional cross section becomes energy dependent at a zero crossing, the three-body recombination loss rate remains energy independent.
In addition we show that $L_e$ continues to be the dominant length in the inelastic three-body processes for larger scattering lengths up to a region where the universal three-body physics takes over and the leading length becomes $a$. 

\begin{figure}
 {\centering \resizebox*{0.48\textwidth}{0.32\textheight}
 {{\includegraphics{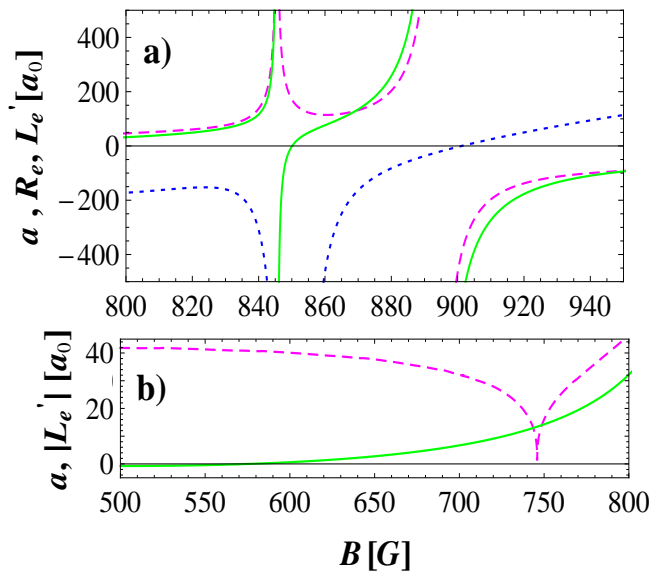}}}
 \par}
\caption{\label{fig:ScatteringLength} Scattering length $a$ (green solid line), effective range $R_e$ (blue dotted line) and effective length $L^\prime_e$ (pink dashed line) in units of Bohr radius $a_0$ as a function of the magnetic field $B$ as obtained from our most up-to-date coupled channels analysis~\citep{Gross11}. (a) - Region of the high field zero crossing with $a=0$ at $850.1$G. $a$ diverges at the positions of Feshbach resonances at $845.5$G and $894$G. $R_e$ diverges when $a\rightarrow 0$. (b) - Region of the low field zero crossing with $a=0$ at $575.9$G. The absolute value of $L^\prime_e$ is shown to emphasize the $L^\prime_e=0$ point at $744.15$G, where $R_e$ (not shown) rapidly changes sign.}
\end{figure}

In the experiment, we use an ultracold sample of $^7$Li atoms prepared in the $|F=1,m_F=0\rangle$ state, where two 
Feshbach resonances allow for the tunability of the scattering length $a$. The first resonance is located at $844.9(8)$G and is closed-channel dominated. The second resonance is located at $893.7(4)$G, and is in between being closed- and open-channel dominated.
These resonances were located and characterized in Ref.~\citep{Gross11} by fitting the binding energies of weakly bound dimers obtained in rf-association spectroscopy with coupled channels calculations. 
Figures~\ref{fig:ScatteringLength}(a) and~\ref{fig:ScatteringLength}(b) show $a$ and $R_e$ dependence on the external magnetic field $B$ from our most up-to-date coupled channels analysis~\citep{Gross11}. 
Divergences of the scattering length correspond to the resonance positions. 
In Ref.~\citep{Gross09}, a log-periodic behavior of the three-body recombination loss rate coefficient has been demonstrated in the vicinity of the $893.7(4)$G resonance in the limit of $a\gg r_{\textrm{vdW}}$. 
Here, in contrast, we are interested in regions where the scattering length crosses zero. In Fig.~\ref{fig:ScatteringLength}, two zero crossings are shown: at $\sim 850$G and $\sim 576$G. 
We ignore the third zero crossing at $\sim 412$G (not shown in the figure) for its apparent similarity to the one at $\sim 576$G from the point of view of the present study. 
As was noted earlier, $R_e$ diverges at zero crossings which can be seen in Fig.~\ref{fig:ScatteringLength}(a). 

First, we consider two-body elastic processes in the vicinity of the $\sim 850$G zero crossing. From Eq.~(\ref{Eq_tandelta}) we obtain the energy-dependent s-wave collisional cross section between two identical bosons
\begin{equation}
\sigma (k)=\frac{8\pi}{k^2} \sin^2(\delta(k))=\frac{8\pi \left(V_ek^2-a\right)^2}{1+\left(V_ek^2-a\right)^2 k^2}.
\label{Eq_Sigma2b}
\end{equation} 
Accordingly, $\sigma(k)$ vanishes when $a=V_ek^2$, reflecting the emerging energy dependence of the two-body collisional cross section. 

\begin{figure}
 {\centering \resizebox*{0.44\textwidth}{0.18\textheight}
 {{\includegraphics{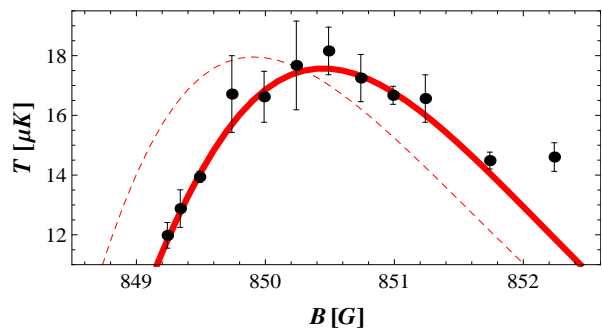}}}
 \par}
\caption{\label{fig:2BodyEalsticS} Temperature at the end of fast evaporation cycle as a function of the magnetic field $B$. Maximal temperature corresponds to minimum in the collisional cross-section. The red solid line is the finite temperature theory model described in text. The fitted position of the scattering length's zero-crossing is $849.92(25)$G. The dashed line represents the model which excludes energy dependence of the cross-section and disagrees with data.}
\end{figure}

Elastic collisions are crucial for evaporative cooling, therefore, the lack of cooling serves as the experimental strategy to find a vanishing cross section $\sigma(k)=0$.
In the experiment, $^7$Li atoms are trapped in a crossed beam optical dipole trap~\citep{Gross08} and we perform a fast cooling at the last stage of the evaporation cycle by lowering the power of the laser beams in 500 ms at different magnetic fields. 
The initial temperature of atoms is measured to be $31.1(3.5)\mu$K and the final temperature, as a function of $B$, is represented in Fig.~\ref{fig:2BodyEalsticS}. 
A maximum in temperature can be clearly identified at $\sim 850.5$G which indicates the total failure of the evaporative cooling routine. 
We model the temperature decrease during evaporation with a set of two coupled rate equations: for number of atoms $N$ and temperature $T$~\citep{supMat}. 
The model includes time evolution of temperature due to evaporation cooling along with adiabatic cooling due to weakening of the optical dipole potential~\cite{O'Hara01}. 
The latter cooling mechanism, being a single-body effect, is independent of $\sigma(k)$ and works equally well for all magnetic field values.
The former becomes inefficient when the elastic collision rate $\Gamma_{el}= n_0 \langle\sigma(k) v\rangle$ is minimized. In this expression, $n_0$ is the peak density of atoms in the trap and $\langle \sigma(k) v\rangle$ implies thermal averaging over the Maxwell-Boltzmann distribution~\citep{supMat}. 

The solid red line in Fig.~\ref{fig:2BodyEalsticS} represents the solution to our model with two fitting parameters: position of the scattering length's zero crossing and the amplitude factor (discussed later) that reflects our experimental uncertainties. 
Except for the last experimental point, the agreement with the data is remarkable which confirms that the two-body elastic collisional cross-section is minimized at the magnetic field of $\sim 850.5$G. 
The position of the scattering length's zero crossing is determined to be at $849.92(25)$G, where the error is dominated by the magnetic field calibration uncertainty.
This value agrees with the coupled channels analysis prediction of $850.1$G (see Fig.~\ref{fig:ScatteringLength}(a)) within less than 1$\sigma$ of the experimental error and suits, even better, the position derived in the most recent coupled channels analysis of collisions in both isotopes of lithium atoms in Ref.~\cite{Julienne14}.
The minimum in the elastic collisional cross section is shifted from the scattering length's zero-crossing position due to the energy dependence of $\sigma(k)$. 
For comparison, the dashed line in the figure represents the same model excluding the energy dependence of the cross-section which disagrees with data.

The same experimental strategy can be applied to identify another $\sigma(k)=0$ position related to the lower magnetic field zero crossing. 
However, weak dependence of $a$ on $B$ in this region allows for a significant tolerance in the uncertainty of the zero-crossing position. 
From the point of view of the present study, this is also indicated by a very weak magnetic field dependence of the three-body effective length $L^\prime_e$ (explained later) around $\sim 576$G (see~Fig.\ref{fig:ScatteringLength}(b)).

Next, we consider the three-body inelastic processes; however, first it 
is worth noting that the $|F=1, m_F=0\rangle$ state is not the absolute ground state of the system, and thus, two-body inelastic dipole-dipole relaxation is allowed. 
However, as shown in Ref.~\citep{Gross11} it is extremely weak for all magnetic field values relevant here, and the dominant trap-loss mechanism is induced by three-body recombination. 

After loading atoms from a magneto-optical trap into an optical dipole trap we perform evaporative cooling at the wing of the narrow resonance. 
Then we jump across the resonance to $858$G within $1$ms and wait for $500$ms before we move within $10$ms to the target field where atom number and temperature time evolutions are measured.
The typical holding time at the final magnetic field varies from $5$ to $20$ sec where we lose $\sim 60\%$ of atoms. 
The three-body recombination loss rate coefficient $K_3$ is extracted from the fit of atom number decay measurements with the solution to the three-body loss rate differential equation: $\dot{N}=-K_{3}\langle n^2\rangle N - \Gamma N$, where $\Gamma$ is the single-body loss rate coefficient due to residual collisions with thermal atoms in vacuum~\citep{supMat}. 
Three-body recombination is known to be accompanied by heating and $K_3$ should be extracted from a set of two coupled rate equations: atom number and temperature time evolutions~\citep{Weber03}. 
However, in the regime of $a\gg r_{vdW}$, even with this precaution $K_3$ usually deviates from the overall $a^4$ scaling by a constant factor. Excluding log-periodic oscillations of $K_3$ due to Efimov physics~\cite{Braaten&Hammer06}, the deviation can be attributed mainly to the systematic uncertainties in the atom number calibration~\cite{Uncertainties}.
An elegant method to calibrate atom number is based on the equation of state measurements recently used to characterize the unitary regime where $|a|\rightarrow \infty$~\citep{Rem13}. 
Here, to calibrate our data we use the amplitude factor for $K_3$ previously determined in our system in the universal regime~\citep{Gross09,Gross10}. 
Analysis of the systematic uncertainties, presented in Supplemental Material, shows that the amplitude factor applied for $K_3$ can be related to the amplitude factor introduced earlier as a fitting parameter in Fig.~\ref{fig:2BodyEalsticS}~\citep{supMat}.
A good agreement between these factors hints at their correct interpretation as being dominated by the experimental uncertainties in the atom number calibration.
Note, also, that using a single rate equation for atom number decay to extract $K_3$ causes only a slight correction to the overall amplitude factor~\citep{supMat}. 

\begin{figure}
 {\centering \resizebox*{0.44\textwidth}{0.18\textheight}
 {{\includegraphics{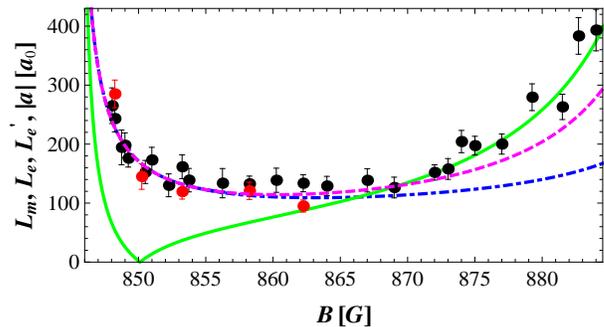}}}
 \par}
\caption{\label{fig:3BodyRecomb} Measurements of the three-body recombination length $L_m$ at $2.5\mu$K (black dots) and $10\mu$K (red dots). Blue dashed-dotted line represents the effective recombination length $L_e$ and pink dashed line shows the modified effective length $L^\prime_e$ that includes $a^3$ term correction (see Eq.~\ref{Eq_EffectiveLength}). Green solid line represents the modulus of the scattering length $|a|$. For comparison, the van der Waals length for $^7$Li atoms is $r_{vdW}=32.5 a_0$.}
\end{figure}

For $a>>r_{\textrm{vdW}}$ the three-body recombination loss rate coefficient is commonly represented as $K_3=3C(a)\hbar a^4/m$. 
The general $a^4$ power dependence is dictated by the resonantly enhanced two-body interactions, while $C(a)$ reflects the discrete scaling invariance of $K_3$ related to Efimov physics~\citep{Braaten&Hammer06}. 
For positive scattering length, $C(a)$ shows log-periodic behavior with the maximum value of $\sim 70$. 
Because of the relatively short lifetime of Efimov trimers in $^7$Li (large inelastic parameter $\eta_*$), this value was measured to be somewhat smaller, $C_{max}\approx 54.7$~\citep{Gross09}. 

The universal $a^4$ dependence is expected to break down when $a\rightarrow 0$. 
Still $K_3$ can be formally represented here as
\begin{equation}
K_3=3C\frac{\hbar}{m}L_m^4,
\label{Eq_3bodyRate}
\end{equation} 
with $L_m$ being a characteristic recombination length for the measured $K_3$ values and $C$ assumed to be a scattering length independent constant. 
Measurements of $L_m$ as a function of the magnetic field $B$ for two different temperatures are shown in Fig.~\ref{fig:3BodyRecomb}, setting $C=C_{max}$, the universal limit's maximal value. 
Above $\sim 865$G we reproduce the measurements from Ref.~\citep{Gross09} where $L_m$ follows $a$. 
For lower values of the magnetic field, deviations from $a$ dependence become evident. 
Instead of decreasing with $a$, $L_m$ saturates at $\sim 120a_0$ where $a_0$ is the Bohr radius and starts to increase again below $\sim 855$G.
Nothing dramatic happens at the point of the scattering length's zero-crossing where $L_m$ continues to increase smoothly.
For comparison, we note that the van der Waals length of $^7$Li is $r_{vdW}=32.5 a_0$.

To analyze the data, we assume that the two-body physics plays a decisive role in three-body recombination loss rates, and that the relevant parameter can be directly extracted from the expansion of the two-body phase shift
\begin{equation}
\delta(k)=-ka+V_ek^3+\frac{k^3a^3}{3}.
\label{Eq_PhaseShift}
\end{equation}

When $a\rightarrow 0$, the dominant term in Eq.~(\ref{Eq_PhaseShift}) is $V_ek^3=(L_ek)^3$ where the effective length is $L_e=(-R_ea^2/2)^{1/3}$. 
$L_e$ is represented in Fig.~\ref{fig:3BodyRecomb} as a blue dashed-dotted line (indistinguishable from the pink line below $855$G). 
Notably, it stays finite for vanishing $a$ and diverging $R_e$. 
Moreover, in the vicinity of $\sim 850$G it agrees remarkably well with the data with no adjustable parameters.
This agreement confirms our assumption that the two-body physics alone defines the behavior of $K_3$.
For larger magnetic field values, $L_e$ starts to slightly deviate from the data.
However, according to Eq.~(\ref{Eq_PhaseShift}), for larger scattering length the two-body scattering volume should be appended by the $a^3$ term and the effective length then becomes
\begin{equation}
L^\prime_e=\left(\frac{a^3}{3}-\frac{R_ea^2}{2}\right)^{1/3}.
\label{Eq_EffectiveLength}
\end{equation}
In Fig.~\ref{fig:3BodyRecomb} $L^\prime_e$ is shown as a pink dashed line and it is in excellent agreement with the data up to $870$G at which the first term in Eq.~(\ref{Eq_PhaseShift}) becomes dominant.

Note that for two-body elastic scattering in the $a\rightarrow 0$ limit, the effective volume $V_e$ is multiplied by $k^2$ (see Eq.~(\ref{Eq_Sigma2b})) to form the effective collisional length which is then explicitly energy dependent. 
On the contrary, $L_e$ and $L^\prime_e$ posses no explicit energy dependence.
As the relevant length for $K_3$ can be constructed differently as compared to Eq.~(\ref{Eq_EffectiveLength}), {\it e.g.} as the aforementioned effective collisional length, the question of energy dependence of $K_3$ remains open.
Thus, we measure the three-body recombination loss rate coefficient at a different temperature, $10\mu$K, and the result is shown in Fig.~\ref{fig:3BodyRecomb} (red dots).
The two measurements are indistinguishable within the experimental errors providing evidence that $K_3$ is energy independent.
As an additional argument, we note that if $L^\prime_e$ is defined according to Eq.~(\ref{Eq_EffectiveLength}), the measured recombination length $L_m$ agrees with it if $C=C_{max}$ in Eq.~(\ref{Eq_3bodyRate}), the same value as in the universal limit where $a>>r_{\textrm{vdW}}$. 
Any attempt to define an energy-dependent effective recombination length, for instance as $(L^\prime_e)^3 k^2$ in analogy to the two-body collisional length, will require tuning of $C$ to a significantly different value.

Finally, we consider the zerocrossing at $\sim 576$G. If we assume, again, $L^{\prime}_e$ to be the relevant length there, $K_3$ can be evaluated from Eq.~(\ref{Eq_3bodyRate}) substituting $L_m$ with $L^{\prime}_e=-40.5a_0$ (see Fig.~\ref{fig:ScatteringLength}(b)). 
This predicts the $K_3$ to be $2$ orders of magnitude smaller than the values measured at $\sim 850$G which is well below our resolution limit for the highest achievable densities set by collisions with residual atoms from vacuum. 
Thus, from our data, we can only extract the upper limit for $L^{\prime}_e$ at $\sim 576$G, $|L^{\prime}_e|<100a_0$, which is consistent with the predicted general trend of being smaller than the one at $\sim 850$G. 

In conclusion, we show a well-behaved three-body recombination loss rate at vanishing two-body scattering length. We identify that the only relevant parameters to define $K_3$ in this limit are $a$ and $R_e$. 
It is interesting to extend our studies to other atomic species. Feshbach resonances in $^7$Li are of intermediate character which is not the case for $^{133}$Cs atoms. 
Therefore, a similar study for $^{133}$Cs, $^{39}$K, and $^{85}$Rb would be interesting to complete the description of $K_3$ at a zero crossing. 
It is also interesting to note that $L^{\prime}_e=0$ at $\sim 744$G (see Fig.~\ref{fig:ScatteringLength}(b)) setting minimum in $K_3$. 
It would be an interesting region for a Bose-Einstein condensate as its lifetime is expected to be maximal there. 
Counter-intuitively, the minimum of three-body losses does not occur at zero scattering length. 
Finally, we note that the recombination length, $L^\prime_e$, may play an important role in the effective range corrections to the lowest energy level in the Efimov spectrum. 

Note that the energy dependence of $K_3$ was reported in the vicinity of a narrow Feshbach resonance in a two-component Fermi gas~\citep{Hazlett12}. 
However, direct three-body recombination in such a mixture is forbidden, and it is mediated through collisions of closed channel molecules with free atoms. 
This mechanism might become relevant in Bose gases for extremely narrow Feshbach resonances.

{\it Note added:} While finalizing this manuscript we became aware of a recently developed theory that supports our findings~\cite{Wang14}.

L. K. and O. M. acknowledge fruitful discussions with C. H. Greene. This research was supported by grants from the United States-Israel Binational Science Foundation (BSF) and the Israel Science Foundation (ISF), Jerusalem, Israel.

\appendix


\section{Model of the evaporation cooling}
The energy evolution in the time dependent optical dipole trap can be expressed as follows~\cite{O'Hara01}:
\begin{equation}
\frac{dE}{dt}=\left(\eta+\tilde{\kappa}\right)k_{b}T\frac{dN}{dt}+\frac{1}{U}\frac{dU}{dt}\frac{E}{2},
\label{eq:EnergyEv}
\end{equation}
where $\eta=\epsilon_c/k_B T(0)$ is the truncation parameter
with $\epsilon_{c}$ being the truncation energy and $\tilde{\kappa}$ stands for the energy carried away by the evaporated atom in addition to $\epsilon_c$. In an harmonic trap, $\tilde{\kappa}$ can be evaluated analytically~\citep{Luiten96}:
\begin{equation}
\tilde{\kappa}=\frac{1-P\left(5,\eta\right)/P\left(3,\eta\right)}{\eta-4 P\left(4,\eta\right)/P\left(3,\eta\right)},
\end{equation}
where $P\left(x,\eta\right)$ is the incomplete $\Gamma$ function. 

The first term in the right hand of Eq.~(\ref{eq:EnergyEv}) describes evaporation cooling. Generally, evaporation in the optical trap is accompanied by weakening of the trapping confinement, thus causing additional adiabatic cooling which is accounted for by the second term in Eq.~(\ref{eq:EnergyEv}). 

The depth of the optical dipole potential is reduced exponentially in time:
\begin{equation}
U(t)=U_{0}\exp\left(-\frac{t}{\tau}\right),
\label{eq:Potential}
\end{equation}
with the time constant $\tau$ defined by the ratio of initial and final potential depths:
\begin{equation}
\tau=\frac{t}{\ln\left(\frac{U_{0}}{U_{f}}\right)}.
\end{equation}
Accordingly, the temporal behaviour of the trap's oscillation frequency is: $\bar{\omega}^{2}\left(t\right)=\bar{\omega}_{0}^{2}\exp{\left(-t/\tau\right)}$
with $\bar{\omega}_0=\left(\omega_{r,0}^{2}\omega_{z,0}\right)^{1/3}$ being the geometric mean of the initial oscillation frequencies.

The atom number loss rate equation due to evaporation is~\citep{Luiten96}:
\begin{equation}
\frac{dN}{dt}=-\Gamma_{ev} N,
\label{eq:AtomNEv}
\end{equation}
where $\Gamma_{ev}=\Gamma_{el}\xi_{ev}$. $\Gamma_{el}=n_{0}\langle\sigma\left(k\right)v\rangle$ is the two-body elastic collision rate. In derivation of Eq.~(\ref{eq:AtomNEv}), $n_0$ is a reference density defined as the central peak density in the limit of $\eta\rightarrow \infty$~\citep{Luiten96}. In an harmonic trap $n_0$ is related to $\bar{\omega}$, $T$ and the total number of atoms $N$: $n_{0}=N\left(\frac{m\bar{\omega}^{2}}{2k_{B}T\pi}\right)^{3/2}$. $\langle\sigma(k)v\rangle$ implies thermal averaging over Maxwell-Boltzmann distribution of the product of two-body collisional cross-section $\sigma\left(k\right)$, defined by Eq. (3) in the main text, and the relative velocity $v=2\hbar k/m$, where factor of $2$ stands for the reduced mass in the center-of-mass coordinate frame:
\begin{equation}
\left\langle \sigma(k)v\right\rangle = \frac{\int_0^\infty \left(2\hbar k/m\right)\sigma\left(k\right)\exp\left(-k^2/k_{th}^2\right)k^2dk}{\int_0^\infty \exp\left(-k^2/k_{th}^2\right)k^2dk},
\label{eq:ThermAvg}
\end{equation}
where $k_{th}^2=(mk_B T)/\hbar^2$. 
The evaporation efficiency $\xi_{ev}=\left(\frac{V_{ev}}{V_{e}}\right)e^{-\eta}$
contains the ratio between the effective volume for elastic collisions leading to evaporation $V_{ev}$ and the single atom volume $V_{e}$~\cite{Luiten96}: 
\begin{equation}
\frac{V_{ev}}{V_{e}}=\eta-4\:\frac{P\left(4,\eta\right)}{P\left(3,\eta\right)}.
\end{equation}

Taking all these factors into account, expanding Eq. (3) of the main text to the lowest order in $k$ and assuming constant $\eta$ we can express Eq.~(\ref{eq:AtomNEv}) in the final form:
\begin{eqnarray}
\frac{dN}{dt}=-\gamma_N \left[\left(-\frac{a}{L_e^2}\frac{1}{\alpha T}+2L_e\right)^2+2L_e^2\right]\times\nonumber\\ \times\left(\alpha T\right) \exp\left(-\frac{3t}{2\tau}\right) N^2,
\label{eq:AtomNEvF}
\end{eqnarray}
where $L_{e}$ is the effective recombination length defined in the main text, and the rate constant of the atom number evolution $\gamma_N$ and $\alpha$ are:
\begin{eqnarray}
\gamma_N=\xi_{ev}\frac{16}{\pi\sqrt{2}}\left(\frac{m \bar{\omega}_0}{\hbar}\right)^2 L_e^4\; \bar{\omega}_0 ,\\
\alpha=\frac{k_{th}^2}{T}=\frac{mk_B}{\hbar^2}.
\end{eqnarray} 

In an harmonic trap the average energy per atom is $E=3k_{b}TN$. Time derivative of this equation, combined with Eqs.~(\ref{eq:EnergyEv}) and (\ref{eq:AtomNEv}), leads to the rate equation for temperature:
\begin{equation}
\frac{dT}{dt}=-\frac{T}{3} \left(\Gamma_{ev}\left(\eta+\tilde{\kappa}-3\right)+ 
\frac{3}{2\tau}\right).
\label{eq:TempEv}
\end{equation}

Finally, integrating Eq.~(\ref{eq:ThermAvg}) and substituting the expression of $\Gamma_{ev}$ to Eq.~(\ref{eq:TempEv}), we obtain:
\begin{eqnarray}
\frac{dT}{dt}=-\gamma_T \left[\left(-\frac{a}{L_e^2}\frac{1}{\alpha T}+2L_e\right)^2+2L_e^2\right]L_e^2\times\nonumber\\
\times \left(\alpha T\right)^2 \left(\frac{\hbar\bar{\omega}_0}{k_B}\right) \exp\left(-\frac{3t}{2\tau}\right) N - \frac{T}{2\tau}\;,
\label{eq:TempEvF}
\end{eqnarray}
where $\gamma_T$ is the rate constant of the temperature evolution:
\begin{equation}
\gamma_T=\xi_{ev}\frac{\left(\eta+\tilde{\kappa}-3\right)}{3}\frac{16}{\pi\sqrt{2}}\left(\frac{m \bar{\omega}_0}{\hbar}\right)L_e^2 \; \bar{\omega}_0.
\end{equation}

Eqs.~(\ref{eq:AtomNEvF}) and (\ref{eq:TempEvF}) form a set of two coupled differential equations describing evaporation cooling dynamics. Their numerical solution is represented in Fig. 2 of the main text.

\section{Measurement of $K_3$}

\begin{figure}
 {\centering \resizebox*{0.44\textwidth}{0.44\textheight}
 {{\includegraphics{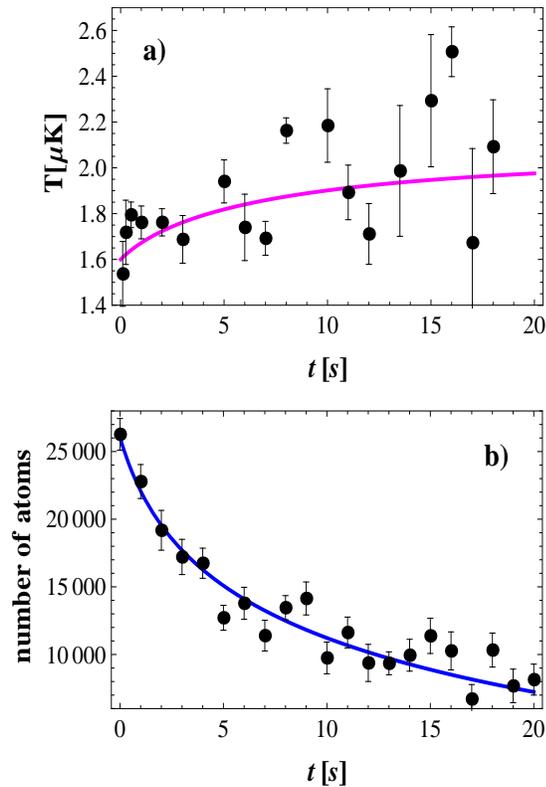}}}
 \par}
\caption{\label{fig:NTEvol} Temperature and atom number time evolution.}
\end{figure}

Three-body recombination rate is locally defined as $K_3 n^3(r)$. The loss rate of atoms is obtained by integrating the three-body recombination rate over the sample's volume: $\dot{N}=-\int K_3 n^3(r)d^3r=-K_3\langle n^2 \rangle N$.
To obtain $K_3$ coefficients we fit the experimentally measured time evolutions of atom number and temperature with two coupled rate equations following the model developed in Ref.~\citep{Weber03}:
\begin{eqnarray}
\frac{dN}{dt}=-\Gamma N - \gamma_3 \frac{N^3}{T^3},\\
\frac{dT}{dt}=\gamma_3 \frac{N^2}{T^3}\frac{T}{3},
\end{eqnarray}
where $\gamma_3=K_3\left(m\omega^2/2\pi k_B\right)^3/\sqrt{27}$ and $\Gamma$ is the single-body loss rate coefficient due to residual collisions with thermal atoms in the vacuum. 
A typical set of measurements at a magnetic field value close to the scattering length zero-crossing is shown in Fig.~\ref{fig:NTEvol}. Combined fit to the rate equations reveals the value of $K_3=6.3(1.9)\times 10^{-27}\;\;cm^6/s$. 
In contrary, if only the atom number decay measurement is fitted keeping $T(t)$ constant, $K_3=4.3(1.3)\times 10^-{27}\;\;cm^6/s$. Thus, fitting the single equation of $N(t)$ causes underestimation of the $K_3$ coefficient by $\sim 46\%$ as far as $60\%$ of atoms are lost. We include this underestimation in the amplitude factor, discussed in the next section, and fit only the atom number loss measurements while keeping the loss fraction constant. 

We note that the initial temperature of atoms is measured every time the loss measurement is performed. We find it to be constant over the whole range of the magnetic field values presented in Fig.~3 of the main text. In addition, oscillation frequencies of the trap are regularly measured to carefully characterize atomic densities.


\section{Calibration of the experimental uncertainties}

In the regime of $a\gg r_{vdW}$, as mentioned in the main text, $K_3$ coefficient usually deviates from the overall $a^4$ scaling by a constant factor.

In order to match this general behaviour we must multiply the $K_3$ coefficient by a factor $\xi^{-2}=3.5^{-2}$ which corrects for errors in the calibration of our experimental parameters. 

To define contributions of the relevant experimentally measured quantities to $\xi$ we recall that by definition given in the previous section, $K_3\varpropto (V/N)^2$, where $V$ is the volume of the sample. In a cylindrically symmetric harmonic trap $V\varpropto T^{3/2}/(\omega_r^2\omega_z$). Thus, $K_3\varpropto T^3/(N^2(\omega_r^2\omega_z)^2)$. 
In this expression temperature is measured by the time-of-flight and thus the experimental uncertainties are buried in the calibration of the size of a camera pixel $(px)$ including magnification of the imaging system, $i.e.\;\;T\varpropto (px)^2$. 
The number of atoms is measured by the absorption shadow of the cloud on the CCD camera. 
It is related to $N\varpropto (px)^2/\sigma_A$, where $\sigma_A$ is the atomic absorption cross-section for weak imaging light of finite linewidth. 
Thus, the calibration error scales as $(px)^2\sigma_A^2/(\omega_z^2\omega_r^4)$ and is corrected by $\xi$. 
The main systematic uncertainty is hidden in $\sigma_A$ and thus in the absolute atom number counting.
Our uncertainties in $\xi^{-2}$ is $\sim 20\%$ dominated by the uncertainties in the oscillation frequencies.

In case of evaporation cooling the finite temperature is defined by $\Gamma_{el}$. In the limit of $a\rightarrow 0$, Eq.~(\ref{eq:TempEvF}) yields $\Gamma_{el}\varpropto (N/V)T^{5/2}$. 
Thus, $\Gamma_{el}\varpropto (K_3)^{-1/2}T^{5/2}$ so that its error scales as $(\xi\cdot\zeta)$ where $\zeta$ implies an additional correction to $\xi$ due to uncertainties in the temperature measurement. 
The experimental error in temperature measurements is well tracked as being $\sim 10\%$. 
This allows for the estimation of $\zeta$ according to $\zeta^{2/5}=1\pm 0.1$. 
Our fitting procedure in Fig.~2 of the main text reveals $(\xi\cdot\zeta) = 2.64(0.27)$ yielding $\zeta=0.75$, in a good agreement with the above estimation. 

In conclusion, we confirm here that the main experimental uncertainty is in the atom number calibration and the amplitude factor $\xi$ accounts for that.


\end{document}